\documentstyle[aps,prl,epsfig]{revtex}
\tightenlines

\newcommand{\be}{\begin{equation}}
\newcommand{\ee}{\end{equation}}

\newcommand{\ve}{\mbox{$\varepsilon$}}

\newcommand{\bmath}{\begin{mathletters}}
\newcommand{\emath}{\end{mathletters}}

\begin{document}

\title{\Large{\bf Interband pairing in extended two- and three-band Hubbard model}}

\vskip0.5cm 

\author{ G. G\'{o}rski, J. Mizia and Krzysztof Kucab}

\address{Institute of Physics, University of Rzesz\'{o}w, ulica Rejtana 16A, \\
35-958 Rzesz\'{o}w, Poland\\}

\maketitle

\vskip0.5cm

\begin{abstract}

\noindent
We study the existence of intraband and interband pairing in the extended two band and three-band Hubbard model. It is shown that including interband pairing significantly 
increases the superconducting critical temperature in comparison with critical temperature of the intraband pairing. This increase allows for decrease of the nearest-neighbor 
interaction constant necessary for critical temperature in high temperature superconductors to the realistic values.

\end{abstract}

\vskip0.5cm



\vskip1.5cm

\noindent {\Large {\bf 1. Introduction} }

\vskip0.5cm

High temperature superconductivity was discovered over fifteen years ago \cite{1}, but still today the mechanism of this phenomena is not uniquely established. The matter is 
made additionally more difficult by complicated band structure of the cuprates. The superconducting state is observed mainly in the $\textrm{CuO}_2$ planes but it is created and 
influenced by the whole structure of cuprates, e.g. the apex oxygen \cite{2,3,4} or other donor of localized level (Resonant Pinning Centers (RPC)  \cite{5}). Even description of 
the single superconducting $\textrm{CuO}_2$ plane is rather complicated since in this plane the relatively lightly bound oxygen holes interact with strongly correlated copper 
holes. In the result we obtain hybridized oxygen and copper holes \cite{6,7}. Interaction between the $\textrm{CuO}_2$ planes and the interaction of copper and oxygen holes 
within the $\textrm{CuO}_2$ plane can be described by the two band model, in which strongly correlated levels are hybridizing with the broad conducting band \cite{4,8,9,10,11}. 
The two band model can be used for superconductivity also for heavy fermion materials \cite{12,13,14}.
In most of the two-band models \cite{4,8,11,12,13} it is assumed that the superconducting gap exist only in one hybridized band. Extension of this concept was presented in 
models in which there were two order parameters \cite{10,15,16}, but the order parameter in the band in which there is no Fermi level has very little influence on the results and 
particularly on the critical temperature \cite{10}. More interesting is the use of additional interband order parameters for superconductivity \cite{17,18}. This concept is similar to 
the concept of interlayer order parameter used in the multilayer models \cite{19,20,21}. 
In this paper we analyze the two-band model using both intraband and interband order parameters of superconductivity. We assume that the broad oxygen band is hybridized 
with the copper level what gives two bands separated with the hybridization gap. In a three-band model we analyze broad band interacting with two correlated copper levels ( 
$3d_{x^2 - y^2 }$ and $3d_{3z^2 - r^2 }$), what gives three hybridized bands. Including the $3d_{3z^2  - r^2 }$ level is stimulated by the experimental reports that there is a several 
percentage content of $3d_{3z^2  - r^2 }$ holes in the electronic spectrum of layered cuprates (see \cite{2,22}).
In this paper we will concentrate mainly on the d-wave superconductivity. Series of experiments by ARPES \cite{23} and SQUID \cite{24} have shown that most of the cuprates 
have d-wave superconductivity. This kind of superconductivity is caused by the negative nearest-neighbor interaction \emph{W}, for which there is no unique explanation until 
now. One proposal was that in the three band model there is a significant difference; $\Delta V = V_z  - V_x $ , where $V_x$($V_z$ ) is the nearest-neighbor interaction between 
copper orbital $3d_{x^2  - y^2 }$($3d_{3z^2  - r^2 }$) and the oxygen band 2\emph{p} \cite{25}. In the strong correlation limit the nearest-neighbor interaction \emph{W} is equal 
to  $\Delta V$ . The problem is arising after we try to fit the value of negative $\Delta V$ to the critical temperature of the order of 100 K. This value for d-wave superconductor 
comes out to be unrealistically high \cite{26}. Yet another source of the negative \emph{W} can be the charge transfer instability \cite{27,28}. Further on in this paper, to create 
the d-wave superconductivity, we assumed the existence of attractive nearest-neighbor interaction \emph{W}, without going into details of its origin. 
In section 2 of this paper we propose the Hamiltonian describing interaction between the broad band (e.g. oxygen band 2\emph{p}) and the localized level (e.g. copper orbital 
3\emph{d}). This Hamiltonian is diagonalised by the standard procedure. To analyze superconductivity in hybridized bands we will use the Hartree-Fock factorization and the 
Green function formalism. In the result we obtain equations for the critical temperature of the d-wave and s-wave superconductor including both; intraband and interband pairing. 
In section 3 we analyze the dependence of critical temperature for d-wave superconductivity and nearest-neighbor interaction constant \emph{W}, on the energy difference 
between center of the broad band and the localized level. In addition we present in this section the influence of the interband pairing on values of critical temperature and 
nearest-neighbor interaction \emph{W} in the three-band model.

\vskip1.0cm 

\noindent {\Large {\bf 2. The Model Hamiltonian}} 

\vskip0.5cm 

The Hamiltonian of the two-band model has the following form
\be
H = H_0  + H_I 
\label{1}
\ee
where the kinetic part $H_0$ is given by

\be
H_0=(\ve_{p}-\mu)\sum\limits_{i\sigma}{n_{p_{i\sigma}}}+t_{pp}\sum\limits_{ij\sigma}{(p_{i\sigma}^+p_{j\sigma}+h.c.)}+(\varepsilon_d-\mu)\sum\limits_{i\sigma}{n_{d_{i
\sigma}}}+t_{pd}\sum\limits_{ij\sigma}{\left({d_{i\sigma}^+p_{j\sigma}+ h.c.}\right)} 
\label{2} 
\ee 
with $p_{i\sigma}^+(p_{i\sigma})$ creating (annihilating) hole in the broad band, $d_{i\sigma }^+(d_{i\sigma})$  creating (annihilating) hole in the localized level, 
$n_{p_{i\sigma}}=p_{i\sigma }^ +  p_{i\sigma }$ is the hole number operator in the broad band, $n_{d_{i\sigma } }  = d_{i\sigma }^ +  d_{i\sigma }$ is the hole number operator 
on the localized level, $t_{pp}$ is the broad band hopping integral, $t_{pd}$ is the integral of hopping between the broad band and the narrow level, $\ve_d$  is the energy of the 
narrow level, $\ve_p$  is the mean energy of the broad band.  

The potential part of the Hamiltonian $H_I$ is given by 

\be
H_I  = U_p \sum\limits_{j\sigma } {n_{p_{j\sigma } } n_{p_{j - \sigma } } }  + U_d \sum\limits_{i\sigma } {n_{d_{i\sigma}}n_{d_{i - \sigma}}}+ 
W\sum\limits_{<ij>\sigma,\sigma '}{n_{d_{i\sigma}}n_{p_{j\sigma '}}} 
, 
\label{3}
\ee		
where $U_p$  is the Coulomb repulsion in the broad band, $U_d$  is the Coulomb repulsion on the narrow level, and $W$ is the attractive nearest-neighbor interaction between 
holes in the broad band and in the narrow level.

We assume strong Coulomb correlation on the localized level ($U_d  \to \infty$), which is taken as the zero energy level ($\ve_d=0$ ). This will change the kinetic part of the 
Hamiltonian to the following form 

\be
H_0  = (\varepsilon _p  - \mu )\sum\limits_{i\sigma } {n_{p_{i\sigma } } }  + t_{pp} \sum\limits_{ij\sigma } {(p_{i\sigma }^ +  } p_{j\sigma }  + h.c.) + \tilde t_{pd} \sum\limits_{ < 
ij > \sigma } {\left( {d_{i\sigma }^ +  p_{j\sigma }  + h.c.} \right)}  - \mu \sum\limits_{i\sigma } {n_{d_{i\sigma } } }, 
\label{4}
\ee
where the normalized hopping integral has the form 

 \be
\tilde t_{pd}   = t_{pd}  Z^{{1 \mathord{\left/
 {\vphantom {1 2}} \right.
 \kern-\nulldelimiterspace} 2}} ,
\label{5}
\ee 									
with factor $Z$ responsible for dynamic (hopping) electron correlation on the localized level. This factor can have one of the following forms 

 \be
Z = 1 - n_d 
\label{6}
\quad\textrm{(mean-field approximation)}	
\ee    	
or

\be
Z = \frac{{1 - n_d }}{{1 - \frac{{n_d }}{2}}}
\label{7}
 \quad\textrm{(according to Gutzwiller approximation \cite{29})}.
\ee 

The kinetic part of Hamiltonian in the momentum space has the form

\be
H_0=\sum\limits_{k\sigma} {(\varepsilon _k-\mu)}p_{k\sigma}^+ p_{k\sigma}+ \sum\limits_{k\sigma} {(\tilde t_{pd,k}d_{k\sigma}^+ p_{k\sigma}}+h.c.)-\mu 
\sum\limits_{k\sigma}{d_{k\sigma}^+ d_{k\sigma}} ,
\label{8}
\ee  
where the dispersion relation for the broad band $\ve_k$, as can be seen from the geometry of the $\textrm{CuO}_2$ plane, is given by

\be
\varepsilon _k  = \varepsilon _p  - 2t_{pp} \left[ {\cos (k_x  + k_y ) + \cos (k_x  - k_y )} \right] .
\label{9}
\ee 

The interband hopping integral in the momentum representation has the form 

\be
\tilde t_{pd,k}   = \tilde t_{pd}  \sum\limits_\delta  {e^{ik\delta } }, 
\label{10}
\ee  		
where $\delta$ is the distance between broad band atom (oxygen) and the nearest narrow level atom (copper).

In the strong correlation limit ($U_d\to \infty$) the second term in Hamiltonian (3) will disappear since there is no double occupancy on the copper atoms. The strong correlation, 
$U_d$, will contribute to the negative nearest-neighbor interaction  $W = V_z  - V_x  < 0$,  when we consider that in reality there are two orbitals on copper ($3d_{x^2  - y^2 }$ and 
$3d_{3z^2  - r^2 }$)  with different strength of interaction $V_x,V_z$  with adjacent oxygen orbitals $2p_{x,y}$  (see [25]). In the result the potential part of the Hamiltonian (3) in 
the momentum representation will take on the form

\be
H_I  = U_p \sum\limits_{kk'\sigma } {p_{k\sigma }^ +  p_{ - k - \sigma }^ +  p_{ - k' - \sigma } p_{k'\sigma } }  + \sum\limits_{kk'\sigma } {W_{kk'} } d_{k\sigma }^ +  p_{ - k - 
\sigma }^ +  p_{ - k' - \sigma } d_{k'\sigma } ,
\label{11}
\ee
where 

\be
W_{kk'}=W\sum\limits_\delta{e^{i\left({k-k'}\right)\delta}}=W(\gamma _k \gamma _{k'}+\eta _k\eta _{k'}) ,
\label{12}
\ee

\be
\gamma _k =\cos k_x + \cos k_y ,
\quad
\eta _k =\cos k_x -\cos k_y .
\label{13}
\ee
 
In pairing potential we neglected the triplet pairing terms.

In this paper we do not analyze in details the source of the negative interaction $W$. We will only mention in here that in addition to the Weber's mechanism there is also a charge 
transfer from copper atoms to apex oxygen atoms laying directly above or beneath it. This charge transfer will create opposite charges on copper atoms and in plane oxygen atoms 
laying next to them, and result in the negative electrostatic $W$. This negative nearest-neighbor interaction will contribute to the d-wave superconductivity in addition to 
mentioned above Weber's mechanism.

To analyze our two-band model first we have to diagonalize the kinetic part of our Hamiltonian given by Eq. (8) using the transformation between original $(p,d)$  and new 
$(\alpha,\beta)$  operators described by

\be
\left( {\begin{array}{*{20}c}
   {p_k^ +  }  \\
   {d_k^ +  }  \\
\end{array}} \right) = \left( {\begin{array}{*{20}c}
   {\frac{{E_k^ +  }}{{\sqrt {(E_k^ +  )^2  + \tilde t_{pd,k}^2 } }}} & {\frac{{E_k^ -  }}{{\sqrt {(E_k^ -  )^2  + \tilde t_{pd,k}^2 } }}}  \\
   {\frac{{\tilde t_{pd,k} }}{{\sqrt {(E_k^ +  )^2  + \tilde t_{pd,k}^2 } }}} & {\frac{{\tilde t_{pd,k} }}{{\sqrt {(E_k^ -  )^2  + \tilde t_{pd,k}^2 } }}}  \\
\end{array}} \right)\left( {\begin{array}{*{20}c}
   {\alpha _k^ +  }  \\
   {\beta _k^ +  }  \\
\end{array}} \right) ,
\label{14}
\ee
where the dispersion relation for hybridized quasiparticles is given by

\be
E_k^ \pm   = \frac{1}{2}\left[ {\varepsilon _k  \pm \sqrt {\varepsilon _k^2  + 4\tilde t_{pd,k}^2 } } \right].
\label{15}
\ee

The kinetic energy of the hybridized state has the form

\be
H_0  = \sum\limits_{k\sigma } {(E_k^ +   - \mu )\alpha _{k\sigma }^ +  \alpha _{k\sigma } }  + \sum\limits_{k\sigma } {(E_k^ -   - \mu )\beta _{k\sigma }^ +  \beta _{k\sigma } } .
\label{16}
\ee

In the next step we transform interaction Hamiltonian (11) using the same transformation of Eq. (14), which diagonalized the kinetic part of the Hamiltonian.  To simplify further 
analysis we will neglect the weak Coulomb correlation in the broad band ($U_p=0$), because this repulsion in mean-field approximation does not influence the d-wave 
superconductivity [30, 31]. In the result we obtain 

\be
\begin{array}{l}
 H_I  = \sum\limits_{kk'\sigma } {W_{kk'}  \left( {\omega _k^ +  \alpha _{k,\sigma }^ +   + \omega _k^ -  \beta _{k,\sigma }^ +  } \right)\left( {u_{ - k}^ +  \alpha _{ - k, - \sigma }^ +   
+ u_{ - k}^ -  \beta _{ - k, - \sigma }^ +  } \right)}  \\ 
 \qquad  \times \left( {u_{ - k'}^ +  \alpha _{ - k', - \sigma }   + u_{ - k'}^ -  \beta _{ - k', - \sigma }  } \right)\left( {\omega _{k'}^ +  \alpha _{k',\sigma }   + \omega _{k'}^ -  \beta 
_{k',\sigma }  } \right) \\ 
 \end{array},
\label{17}
\ee
where

\be
u_k^ \pm   = \frac{{E_k^ \pm  }}{{\sqrt {(E_k^ \pm  )^2  + \tilde t_{pd,k}^2 } }}
\quad\textrm{and}\quad
\omega_k^ \pm   = \frac{{\tilde t_{pd,k}  }}{{\sqrt {(E_k^ \pm  )^2  + \tilde t_{pd,k}^2 } }}.
\label{18}
\ee
 
In Hamiltonianu (17) we have 16 combinations of operators $\alpha$ and $\beta$. For each of them we use the Hartree-Fock factorization 

\be
\begin{array}{l}
 \omega \alpha _{k,\sigma }^ +  \alpha _{ - k, - \sigma }^ +  \alpha _{ - k', - \sigma } \alpha _{k',\sigma }  =  < \alpha _{k,\sigma }^ +  \alpha _{ - k, - \sigma }^ +   > \alpha _{ - k', - 
\sigma } \alpha _{k',\sigma }  \\ 
 \qquad  \qquad  + \alpha _{k,\sigma }^ +  \alpha _{ - k, - \sigma }^ +   < \alpha _{ - k', - \sigma } \alpha _{k',\sigma }  >  +  < \alpha _{k,\sigma }^ +  \alpha _{ - k, - \sigma }^ +   >  < 
\alpha _{ - k', - \sigma } \alpha _{k',\sigma }  >  \\ 
 \end{array}.
\label{19}
\ee

In the result, after omitting the constant terms (the last one in Eq. (19)), we obtain from Hamiltonian (17) the following equation

\be
H_I  = \sum\limits_{k,\sigma } {\left( {\begin{array}{*{20}c}
   {\alpha _{k,\sigma } }  \\
   {\alpha _{ - k, - \sigma }^ +  }  \\
   {\beta _{k,\sigma } }  \\
   {\beta _{ - k, - \sigma }^ +  }  \\
\end{array}} \right)^ +  } \left[ {\begin{array}{*{20}c}
   0 & {f_{k,1} \Delta _k^* } & 0 & {f_{k,3} \Delta _k^* }  \\
   {f_{k,1} \Delta _k } & 0 & {f_{k,4} \Delta _k } & 0  \\
   0 & {f_{k,4} \Delta _k^* } & 0 & {f_{k,2} \Delta _k^* }  \\
   {f_{k,3} \Delta _k } & 0 & {f_{k,2} \Delta _k } & 0  \\
\end{array}} \right]\left( {\begin{array}{*{20}c}
   {\alpha _{k,\sigma } }  \\
   {\alpha _{ - k, - \sigma }^ +  }  \\
   {\beta _{k,\sigma } }  \\
   {\beta _{ - k, - \sigma }^ +  }  \\
\end{array}} \right),
\label{20}
\ee
where the functions $f_{k,i}$  are defined as

\be
f_{k,1}=\omega _k^+ u_k^+ ,
f_{k,2}  = \omega _k^-  u_k^- ,
f_{k,3}  = \omega _k^ +  u_k^ - ,
f_{k,4}  = \omega _k^ -  u_k^ + .
\label{21}
\ee

The superconducting order parameter $\Delta _k$  is expressed as

\be
\Delta _k =\Delta _{k,1}+\Delta _{k,2}+\Delta _{k,3}+\Delta _{k,4} ,
\label{22}
\ee
where

\be
\Delta _{k,1}  = \sum\limits_{k'} {W_{kk'} f_{k',1}  < \alpha _{k',\sigma }^ +  \alpha _{ - k', - \sigma }^ +   > },
\label{23}
\ee

\be
\Delta _{k,2}  = \sum\limits_{k'} {W_{kk'} f_{k',2}  < \beta _{k',\sigma }^ +  \beta _{ - k', - \sigma }^ +   > } ,
\label{24}
\ee

\be
\Delta _{k,3}  = \sum\limits_{k'} {W_{kk'} f_{k',3}  < \alpha _{k',\sigma }^ +  \beta _{ - k', - \sigma }^ +   > },
\label{25}
\ee

\be
\Delta _{k,4}  = \sum\limits_{k'} {W_{kk'} f_{k',4}  < \beta _{k',\sigma }^ +  \alpha _{ - k', - \sigma }^ +   > }.
\label{26}
\ee

This order parameter is the sum of intraband parameters ($\Delta _{k,2}$ and $\Delta _{k,2}$), and interband parameters ($\Delta _{k,3}$ and $\Delta _{k,4}$). 
Using Hamiltonian given by the sum of Eqs (16) and (20) in the equations of motion for Green functions 

\be
\varepsilon \left\langle {\left\langle {A;B} \right\rangle } \right\rangle _\varepsilon   = \left\langle {\left[ {A,B} \right]_ +  } \right\rangle  + \left\langle {\left\langle {\left[ {A,H} 
\right]_{ - } ;B} \right\rangle } \right\rangle _\varepsilon ,
\label{27}
\ee
we obtain the following results 

\be
\left[ {\begin{array}{*{20}c}
   {\varepsilon  - E_k^ +   + \mu } & {f_{k,1} \Delta _k } & 0 & {f_{k,4} \Delta _k }  \\
   {f_{k,1} \Delta _k^* } & {\varepsilon  + E_k^ +   - \mu } & {f_{k,3} \Delta _k^* } & 0  \\
   0 & {f_{k,3} \Delta _k } & {\varepsilon  - E_k^ -   + \mu } & {f_{k,2} \Delta _k }  \\
   {f_{k,4} \Delta _k^* } & 0 & {f_{k,2} \Delta _k^* } & {\varepsilon  + E_k^ -   - \mu }  \\
\end{array}} \right]\hat G(k,\varepsilon ) = \left[ {\begin{array}{*{20}c}
   1 & 0 & 0 & 0  \\
   0 & 1 & 0 & 0  \\
   0 & 0 & 1 & 0  \\
   0 & 0 & 0 & 1  \\
\end{array}} \right] .
\label{28}
\ee
 
The solution of Eqs (28) is the Green function given by the matrix

\be
\hat G(k,\varepsilon ) = \left[ {\begin{array}{*{20}c}
   {\varepsilon  - E_k^ +   + \mu } & {f_{k,1} \Delta _k^{} } & 0 & {f_{k,4} \Delta _k^{} }  \\
   {f_{k,1} \Delta _k^* } & {\varepsilon  + E_k^ +   - \mu } & {f_{k,3} \Delta _k^* } & 0  \\
   0 & {f_{k,3} \Delta _k^{} } & {\varepsilon  - E_k^ -   + \mu } & {f_{k,2} \Delta _k^{} }  \\
   {f_{k,4} \Delta _k^* } & 0 & {f_{k,2} \Delta _k^* } & {\varepsilon  + E_k^ -   - \mu }  \\
\end{array}} \right]^{ - 1} ,
\label{29}
\ee
where for example 

\be
G_{12} (k,\varepsilon ) = \frac{{f_{k,1} \Delta _k \left[ { - \varepsilon ^2  + (E_k^ -   - \mu )^2 } \right]}}{{D(\varepsilon )}}  ,
\label{30}
\ee
 
\be
G_{34} (k,\varepsilon ) = \frac{{f_{k,2} \Delta _k \left[ { - \varepsilon ^2  + (E_k^ +   - \mu )^2 } \right]}}{{D(\varepsilon )}} ,
\label{31}
\ee
 
\be
G_{14} (k,\varepsilon ) = \frac{{f_{k,4} \Delta _k \left[ { - (\varepsilon  - E_k^ -   + \mu )(\varepsilon  + E_k^ +   - \mu )} \right]}}{{D(\varepsilon )}} ,
\label{32}
\ee
 
\be
G_{32} (k,\varepsilon ) = \frac{{f_{k,3} \Delta _k \left[ { - (\varepsilon  + E_k^ -   - \mu )(\varepsilon  - E_k^ +   + \mu )} \right]}}{{D(\varepsilon )}} .
\label{33}
\ee

Denominator, $D(\varepsilon)$, for all these functions is given by 

\be
\begin{array}{l}
 D(\varepsilon ) = \left[ {f_{k,1}^2 \Delta _k^2  - \varepsilon ^2  + (E_k^ +   - \mu )^2 } \right]\left[ {f_{k,2}^2 \Delta _k^2  - \varepsilon ^2  + (E_k^ -   - \mu )^2 } \right] \\ 
 \qquad \quad + \left[ {f_{k,3}^2 \Delta _k^2  - (\varepsilon  - E_k^ -   + \mu )(\varepsilon  + E_k^ +   - \mu )} \right]\left[ {f_{k,4}^2 \Delta _k^2  - (\varepsilon  + E_k^ -   - \mu 
)(\varepsilon  - E_k^ +   + \mu )} \right] \\ 
\qquad \quad  - 2f_{k,1} f_{k,2} f_{k,3} f_{k,4} \Delta _k^4  - \left[ {\varepsilon ^2  - (E_k^ +   - \mu )^2 } \right]\left[ {\varepsilon ^2  - (E_k^ -   - \mu )^2 } \right] \\ 
 \end{array} .
\label{34}
\ee

Replacing in formulae (23)-(26) the mean values by the Green functions we obtain 

\be
\Delta _{k,1}  =  - \sum\limits_{k'} {W_{kk'}^{} f_{k',1} \int {\frac{{f(\varepsilon )}}{\pi }{\mathop{\rm Im}\nolimits} G_{12} (\varepsilon ,k'} )d\varepsilon } ,
\label{35}
\ee

\be
\Delta _{k,2}  =  - \sum\limits_{k'} {W_{kk'}^{} f_{k',2} \int {\frac{{f(\varepsilon )}}{\pi }{\mathop{\rm Im}\nolimits} G_{34} (\varepsilon ,k'} )d\varepsilon } ,
\label{36}
\ee

\be
\Delta _{k,3}  =  - \sum\limits_{k'} {W_{kk'}^{} f_{k',3} \int {\frac{{f(\varepsilon )}}{\pi }{\mathop{\rm Im}\nolimits} G_{32} (\varepsilon ,k'} )d\varepsilon } ,
\label{37}
\ee
 
\be
\Delta _{k,4}  =  - \sum\limits_{k'} {W_{kk'}^{} f_{k',4} \int {\frac{{f(\varepsilon )}}{\pi }{\mathop{\rm Im}\nolimits} G_{14} (\varepsilon ,k'} )d\varepsilon } .
\label{38}
\ee
 
The total superconducting gap $\Delta _k$  is given by 

\be
\Delta _k  =  - \sum\limits_{k'} {W_{kk'}^{} \int {\frac{{f(\varepsilon )}}{\pi }{\mathop{\rm Im}\nolimits} \left[ {f_{k',1} G_{12} (\varepsilon ,k') + f_{k',2} G_{34} (\varepsilon 
,k')} \right.} } \left. { + f_{k',3} G_{32} (\varepsilon ,k') + f_{k',4} G_{14} (\varepsilon ,k')} \right]d\varepsilon .
\label{39}
\ee

The symmetry of the gap $\Delta _k$ depends on the symmetry of the pairing potential $W_{kk'}$. From Eq. (12) one can see that the order parameter has to be written as

\be
\Delta _k  = \Delta _s \gamma _k  + \Delta _d \eta _k .
\label{40}
\ee

Significant advantage of the current method lies in the fact that the superconducting state has pure d-wave and s-wave 
symmetry. In the two bands models with only intraband ordering  [11-13] symmetry of the order parameter depended also on the hybridization factor (e.g.  $g_k$ in [12,13]). The 
present definition of the order parameter (Eq. (40)) has the standard form for the s+d wave superconductivity.

Inserting to Eq. (39) both the superconducting gap (40) and the effective potential (12), we obtain two separate equations describing two types of superconductivity 

\noindent - for the s-wave superconductivity

\be
1 =  - W\sum\limits_{k'} {\gamma _{k'}^2 \int {\frac{{f(\varepsilon )}}{\pi }{\mathop{\rm Im}\nolimits} \left[ {f_{k',1} G_{12}^* (\varepsilon ,k') + f_{k',2} G_{34}^* (\varepsilon 
,k')} \right.} } \left. { + f_{k',3} G_{32}^* (\varepsilon ,k') + f_{k',4} G_{14}^* (\varepsilon ,k')} \right]d\varepsilon ,
\label{41}
\ee
- for the d-wave superconductivity

\be
1 =  - W\sum\limits_{k'} {\eta _{k'}^2 \int {\frac{{f(\varepsilon )}}{\pi }{\mathop{\rm Im}\nolimits} \left[ {f_{k',1} G_{12}^* (\varepsilon ,k') + f_{k',2} G_{34}^* (\varepsilon ,k')} 
\right.} } \left. { + f_{k',3} G_{32}^* (\varepsilon ,k') + f_{k',4} G_{14}^* (\varepsilon ,k')} \right]d\varepsilon ,
\label{42}
\ee
where

\be
G_{ij}^* (\varepsilon ,k) = \frac{{G_{ij} (\varepsilon ,k)}}{{\Delta _k }} .
\label{43}
\ee
 
The Fermi level used in Eqs (41) and (42) is defined by the condition for carrier concentration in the narrow band 

\be
n_{d\sigma }  = \sum\limits_k {\int {\frac{{f(\varepsilon )}}{\pi }{\mathop{\rm Im}\nolimits} \left\{ {(\omega _k^ +  )^2 G_{11}^{} (\varepsilon ,k') + \omega _k^ +  \omega _k^ -  
\left[ {G_{13}^{} (\varepsilon ,k) + G_{31}^{} (\varepsilon ,k)} \right]} \right.} }  + \left. {(\omega _k^ -  )^2 G_{33}^{} (\varepsilon ,k)} \right\}d\varepsilon .
\label{44}
\ee

Analytical solution of Eqs (41) and (42) for the nonzero gap ($\Delta _k\ne 0$, $T<T_C$) is very complicated. In the critical temperature; $T \to T_C $, $\Delta _k  \to 0$, and the 
solution of Eqs (41) and (42) simplifies to the following forms 

\noindent - for the s-wave superconductivity

\be
\begin{array}{l}
  1 =  - W\sum\limits_{k'} {\gamma _{k'}^2 \left[ {\frac{{f_{k',1}^2 }}{{2(E_{k'}^ +   - \mu )}}\tanh \left( {\frac{\beta }{2}(E_{k'}^ +   - \mu )} \right) + \frac{{f_{k',2}^2 
}}{{2(E_{k'}^ -   - \mu )}}\tanh \left( {\frac{\beta }{2}(E_{k'}^ -   - \mu )} \right)} \right.}  \\
\qquad
  {\rm           }\left. { + \frac{{f_{k',3}^2  + f_{k',4}^2 }}{{2(E_{k'}^ +   + E_{k'}^ -   - 2\mu )}}\left( {\tanh \left( {\frac{\beta }{2}(E_{k'}^ +   - \mu )} \right) + \tanh \left( {\frac{\beta 
}{2}(E_{k'}^ -   - \mu )} \right)} \right)} \right] \\
\end{array} ,
\label{45}
\ee
- for the d-wave superconductivity

\be
\begin{array}{l}
  1 =  - W\sum\limits_{k'} {\eta _{k'}^2 \left[ {\frac{{f_{k',1}^2 }}{{2(E_{k'}^ +   - \mu )}}\tanh \left( {\frac{\beta }{2}(E_{k'}^ +   - \mu )} \right) + \frac{{f_{k',2}^2 }}{{2(E_{k'}^ -   
- \mu )}}} \right.} \tanh \left( {\frac{\beta }{2}(E_{k'}^ -   - \mu )} \right) \\
\qquad 
  {\rm           }\left. { + \frac{{f_{k',3}^2  + f_{k',4}^2 }}{{2(E_{k'}^ +   + E_{k'}^ -   - 2\mu )}}\left( {\tanh \left( {\frac{\beta }{2}(E_{k'}^ +   - \mu )} \right) + \tanh \left( {\frac{\beta 
}{2}(E_{k'}^ -   - \mu )} \right)} \right)} \right] \\
\end{array} .
\label{46}
\ee
 
The above two equations together with Eq. (44) at $\Delta _k  \to 0$ are allowing for calculating the dependence of critical temperature on carrier concentration, which will be 
analyzed in the next section. 
 
 \vskip1.0cm

\noindent {\Large {\bf 3. Numerical results and discussion}}

\vskip0.5cm

To compare Eq. (46) with analogous equation for the two-band model of superconductivity with order gap only in one band [11-13] we insert limiting values of $f_{k,1} 
=f_{k,3}=f_{k,4}=0$  and $f_{k,2}  \ne 0$ into Eq. (46), obtaining the following equation

\be
1 =-W\sum\limits_{k'}{\eta _{k'}^2\frac{{\left({E_{k'}^- \tilde t_{pd,k'}}\right)^2}}{{\left({(E_{k'}^-)^2 +\tilde t_{pd,k'}^2}\right)^2}}\frac{{\tanh \left({\frac{\beta }{2}(E_{k'}^- 
-\mu)}\right)}}{{2(E_{k'}^- -\mu)}}} ,
\label{47}
\ee
where $f_{k,2}={{E_k^- \tilde t_{pd,k'}}\mathord{\left/{\vphantom{{E_k^- \tilde t_{pd,k'}}{\left[{(E_k^-)^2 +\tilde t_{pd,k'}^2}\right]}}}\right.\kern-\nulldelimiterspace} {\left[ 
{(E_k^ -  )^2  + \tilde t_{pd,k'}^2 } \right]}}$. 

Including terms with  $f_{k,1}$, $f_{k,3}$  and $f_{k,4}$ in Eq. (46) allows for obtaining the critical temperature of 100 K at smaller values of nearest-neighbor interaction. Equation 
(47) gives this temperature at $\ve= 2t_{pd}$, $t_{pp}=0.15t_{pd}$, $t_{pd}=1{\rm eV}$ and $n_d=0.75$ for the nearest-neighbor interaction $W=-0.7t_{pd}$, while from Eq. (46) 
the same temperature at the same band structure parameters can be reached already at $W=-0.6t_{pd}$. This gives 15\% reduction of the necessary interaction $W$. The main 
contribution to the reduction of $W$ comes from the term with $f_{k,4}$, which represents the interband pairing (Eq. (20)). The other two terms (with $f_{k,1}$ and $f_{k,3}$) 
have only small influence on $W$, e.g. the $f_{k,1}$ term, which is responsible for pairing in the upper sub-band (see Eq. (20)) decreases the value of the nearest-neighbor 
interaction $W$ for only 1\%. For this reason the upper sub-band was neglected in the previous papers with only intraband order parameter left [11-13]. For simplicity we assume 
that the interband hopping integral does not depend on the wave vector $k$ ($\tilde t_{pd,k}\equiv \tilde t_{pd}$).

\begin{figure}
\begin{center}
\epsfig{file=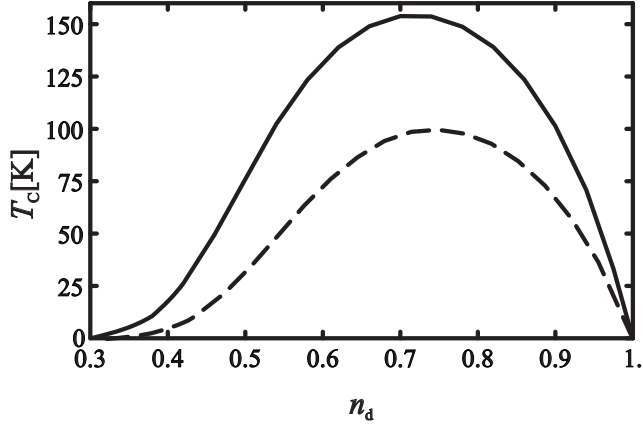,width=0.47\hsize}
\end{center}
\vskip0.5cm
\caption{Critical temperature $T_C$ for the d-wave pairing as a function of the carrier concentration in the narrow band $n_d$ estimated from Eq. (46) (solid line) and Eq. (47) 
(dashed line), at $\ve=2t_{pd}$, $t_{pp}=0.15t_{pd}$, $t_{pd}=1{\rm eV}$, $W=-0.6t_{pd}$.}
\vskip0.5cm
\end{figure}

At the same value of interaction $W$ the model with interband pairing (Eq. (46)) will give larger values of the critical temperature than the model with only intraband pairing (Eq. 
(47)). Fig. 1 presents the critical temperature dependence on carrier concentration of copper estimated from Eqs (46) and (47). Both these curves were calculated for the same set of 
parameters; $\ve=2t_{pd}$, $t_{pp}=0.15t_{pd}$, $t_{pd}=1{\rm eV}$ and $W=-0.679t_{pd}$. One can see that the maximum of critical temperature is rising for about 50\% when 
pairing takes place not only in one band. 

\begin{figure}
\begin{center}
\epsfig{file=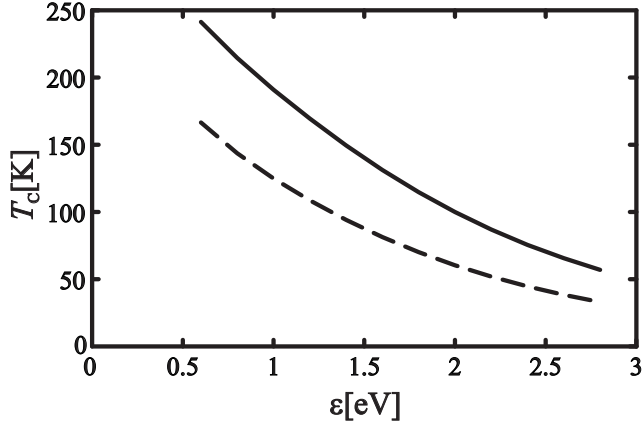,width=0.47\hsize}
\end{center}
\vskip0.5cm
\caption{Critical temperature $T_C$ for the d-wave pairing as a function of energy difference between center of the broad band and the narrow level $\ve$ with interband pairing 
(solid line) and without it (dashed line), at $t_{pp}=0.15t_{pd}$, $t_{pd}=1{\rm eV}$, $W=-0.6t_{pd}$ and $n_d= 0.75$}
\vskip0.5cm
\end{figure}

The influence of interband pairing on maximum of critical temperature and value of necessary nearest-neighbor interaction grows with decreasing energy difference between the 
localized level and the center of the broad band. Fig. 2 shows dependence of the critical temperature on this energy difference $\ve$, with interband pairing (solid line) and without 
it (dashed line), at  $t_{pp}=0.15t_{pd}$, $t_{pd}=1{\rm eV}$, $W=-0.6t_{pd}$ and $n_d= 0.75$. Dependence of the nearest-neighbor interaction necessary for reaching 
$T_C=100\rm{K}$ on the energy difference $\ve$ is given in Fig. 3 for $t_{pp}=0.15t_{pd}$, $t_{pd}=1{\rm eV}$, $n_d= 0.75$. Decreasing energy gap of hybridized bands; 
$E_k^+ - E_k^-$ with decreasing $\ve$ causes increase of the critical temperature and decrease of interaction $W$. 

\begin{figure}
\begin{center}
\epsfig{file=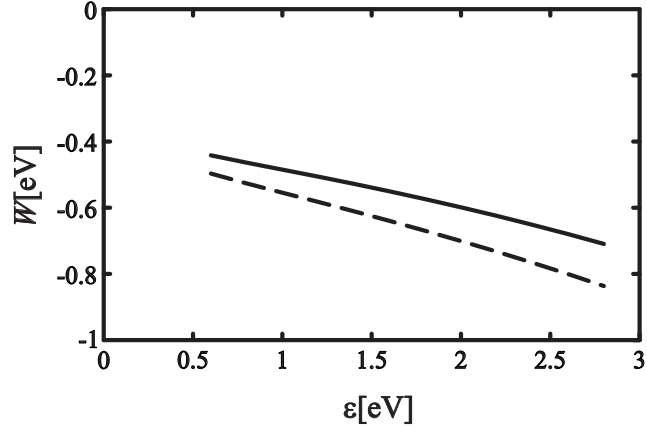,width=0.47\hsize}
\end{center}
\vskip0.5cm
\caption{Nearest-neighbor interaction $W$ as a function of energy difference $\ve$ with interband pairing (solid line) and without it (dashed line), at $t_{pp}=0.15t_{pd}$, 
$t_{pd}=1{\rm eV}$, $n_d= 0.75$and $T_C=100\rm{K}$. }
\vskip0.5cm
\end{figure}

The above analysis will be even more important in the three band model describing the interaction between degenerated oxygen orbital O (2p) and two copper orbitals ($3d_{x^2  - 
y^2 }$ and $3d_{3z^2  - r^2 }$), see G\'{o}rski et. al. [26]. The kinetic part of this model is given by

\be
\begin{array}{l}
H_0  = (\varepsilon _p  - \mu )\sum\limits_{k,\sigma } {p_{k\sigma }^ +  p_{k\sigma } }  + (\varepsilon _x  - \mu )\sum\limits_{k,\sigma } {d_{xk\sigma }^ +  d_{xk\sigma } }  + 
(\varepsilon _z  - \mu )\sum\limits_{k,\sigma } {d_{zk\sigma }^ +  d_{zk\sigma } }  \\
\quad \quad
- \sum\limits_{k,\sigma } {\left( {it_{xk} d_{xk\sigma }^ +  p_{k\sigma }  + h.c.} \right)}  + \sum\limits_{k,\sigma } {\left( {it_{zk} d_{zk\sigma }^ +  p_{k\sigma }  + h.c.} \right)} \\
\end{array},
\label{48}
\ee
where

\be
t_{xk}  = 2\tilde t_{pd} \left( {\sin \frac{{k_x a}}{2} - \sin \frac{{k_y a}}{2}} \right) ,
t_{zk}  = \frac{{2\tilde t_{pd} }}{{\sqrt 3 }}\left( {\sin \frac{{k_x a}}{2} + \sin \frac{{k_y a}}{2}} \right) ,
\label{49}
\ee
$d_{xk\sigma}^+ (d_{zk\sigma}^+)$ - are operators creating hole on the $3d_{x^2-y^2}$($3d_{3z^2-r^2}$) level.

The interaction part of Hamiltonian is given by Eq. (11) with $d_{k\sigma}^+ (d_{k\sigma})$ being now the creation (annihilation) hole operators on the $3d_{3z^2-r^2}$ level of 
copper.

The effective pairing potential used by us previously (Eq. (14) in [26]) was only one from the 36 pairing terms coming from diagonalization of the model Hamiltonian. Although the 
chosen single term is dominating superconducting pairing, but as it was shown above, the other terms can be important as well, particularly terms with operators $\chi_k$, where 
there is a small energy difference between lower hybridized band $\beta_k$ and the middle band $\chi_k$. 
Equation describing order parameter and critical temperature of the d-wave superconductor in this the three-band model will have six terms and has the following form

\be
1 =  - W\sum\limits_{k'} {\eta _{k'}^2 \left[ {C_{k',1}  + C_{k',2}  + C_{k',3}  + C_{k',5}  + C_{k',5}  + C_{k',6} } \right]} , 
\label{50}
\ee
where coefficients $C_{k',i}$ are given in Appendix. 

Numerical analysis of Eq. (50) shows that the nearest-neighbor interaction creating $T_C^d\approx 100\rm{K}$ without interband pairing at $\ve=2t_{pd}$, $t_{pp}=0$, 
$n_d=0.81$ is equal to $W=-11.04t_{pd}$. With the interband pairing, at the same structural parameters, this interaction drops to $W=-4.94t_{pd}$. Replacing the oxygen level in 
the three band model by the oxygen band $\ve_k=\ve_p-2t_{pp}\left[{\cos(\frac{{k_x+k_y}}{2})-\cos(\frac{{k_x-k_y}}{2})}\right]$ with $t_{pp}=0.5t_{pd}$ (still 
$\ve=2t_{pd}$ and $n_d=0.75$) reduces the interaction to $W=-8.02t_{pd}$ (see [26]). Including additionally interband pairing reduces this value further to $W=-4.33t_{pd}$. 
Concluding; both effects of oxygen band broadening and interband pairing considered together allow for achieving small and realistic values of the nearest-neighbor 
charge-charge interaction.

\vskip1.0cm 

\noindent {\Large {\bf Acknowledgement}}
\vskip0.5cm

We would like to express our gratitude to Prof. Karol I. Wysoki\'{n}ski from Marie Curie-Sklodowska University for suggesting the problem and for helpful discussions during the 
course of this work.

\vskip1.0cm 

\noindent {\Large {\bf Appendix}}
\vskip0.5cm 

Coefficients $C_{k',i}$ used in Eq. (50) are defined as follows

$$
C_{k',1}=\frac{{(\tilde t_{pd}E_{k'}^+)^2}}{{3\left({(E_{k'}^+)^2 +\Omega _{k'}^2}\right)^2}}\frac{{\tanh \left( {\frac{\beta}{2}(E_{k'}^+ -\mu)}\right)}}{{2(E_{k'}^+ -\mu)}} ,
$$ 				

\[
C_{k',2}=\frac{{(\tilde t_{pd} E_{k'}^-)^2}}{{3\left({(E_{k'}^-)^2 +\Omega _{k'}^2}\right)^2}}\frac{{\tanh \left({\frac{\beta}{2}(E_{k'}^- -\mu)}\right)}}{{2(E_{k'}^- -\mu)}} ,
\]

\[
C_{k',3}=\frac{{(\tilde t_{pd} E_{k'}^-)^2}}{{3\left({(E_{k'}^-)^2 +\Omega _{k'}^2} \right)\left({(E_{k'}^+)^2 +\Omega _{k'}^2} \right)}}\frac{{\left({\tanh 
\left({\frac{\beta}{2}(E_{k'}^+ -\mu)} \right)+\tanh \left({\frac{\beta}{2}(E_{k'}^- -\mu)}\right)}\right)}}{{2(E_{k'}^+ +E_{k'}^- -2\mu)}} ,
\]

\[
C_{k',4}=\frac{{(\tilde t_{pd}E_{k'}^+)^2}}{{3\left({(E_{k'}^-)^2 +\Omega _{k'}^2}\right)\left({(E_{k'}^+ )^2 +\Omega _{k'}^2}\right)}}\frac{{\left({\tanh 
\left({\frac{\beta}{2}(E_{k'}^+ -\mu)} \right)+\tanh \left({\frac{\beta}{2}(E_{k'}^- -\mu)}\right)}\right)}}{{2(E_{k'}^+ +E_{k'}^- -2\mu)}} ,
\]

\[
C_{k',5}=\frac{{(\tilde t_{pd}E_{k'}^+)^2}}{{\Omega _{k'}^2\left({(E_{k'}^+)^2 +\Omega _{k'}^2} \right)}}\frac{{\left({\tanh \left({\frac{\beta }{2}(E_{k'}^+ -\mu)}\right)+\tanh 
\left({\frac{\beta}{2}(E_{k'}^0 -\mu)}\right)}\right)}}{{2(E_{k'}^+ +E_{k'}^0 -2\mu)}} ,
\]

 \[
C_{k',6}=\frac{{(\tilde t_{pd}E_{k'}^-)^2}}{{\Omega _{k'}^2\left({(E_{k'}^-)^2 +\Omega _{k'}^2}\right)}}\frac{{\left( {\tanh \left({\frac{\beta}{2}(E_{k'}^- -\mu)}\right)+\tanh 
\left({\frac{\beta}{2}(E_{k'}^0-\mu)}\right)} \right)}}{{2(E_{k'}^- +E_{k'}^0 -2\mu)}} ,
\]

where 

\[
E_k^ \pm   = {1 \mathord{\left/
 {\vphantom {1 2}} \right.
 \kern-\nulldelimiterspace} 2}\left[ {\varepsilon  \pm \sqrt {\varepsilon ^2  + 4\Omega _k^2 } } \right] ,
\]
  					
\[
E_k^0  = 0
\]

\[
\Omega _k^2  = 4\tilde t_{pd}^2 \left( {\sin \frac{{k_x a}}{2} - \sin \frac{{k_y a}}{2}} \right)^2  + \frac{{4\tilde t_{pd}^2 }}{3}\left( {\sin \frac{{k_x a}}{2} + \sin \frac{{k_y 
a}}{2}} \right)^2 .
\]

\end{document}